\documentclass[twocolumn,aps,prl,showpacs,floatfix,nofootinbib]{revtex4}

\usepackage{amssymb,amsmath}
\usepackage{epsfig}

\newcommand{\be}{\begin{equation}}
\newcommand{\ee}{\end{equation}}

\begin{document}

\title{Nuclear quantum optics with x-ray laser pulses} 

\author{Thomas~J.~\surname{B\"urvenich}}
\email{buervenich@fias.uni-frankfurt.de}
\altaffiliation[New address: ]{Frankfurt Institute for Advanced Studies, Max-von-Laue-Str. 1,  60438 Frankfurt,
Germany}

\author{J\"org~\surname{Evers}}
\email{joerg.evers@mpi-hd.mpg.de}

\author{Christoph~H.~\surname{Keitel}}
\email{keitel@mpi-hd.mpg.de}

\affiliation{Max--Planck--Institut f\"ur Kernphysik,
Saupfercheckweg 1, 69117 Heidelberg, Germany}

\date{\today}

\begin{abstract}
The direct interaction of nuclei with super-intense laser fields is studied.
We show that present and upcoming high-frequency laser facilities, especially
together with a moderate acceleration of the target nuclei to
match photon and transition frequency, do allow for
resonant laser-nucleus interaction. These direct interactions may be 
utilized for the model-independent optical measurement of nuclear 
properties such as the transition frequency and the 
dipole moment, thus opening the field of nuclear quantum 
optics.
As ultimate goal, one may hope that  direct laser-nucleus
interactions could become a versatile tool to enhance 
preparation, control and detection in nuclear physics.
\end{abstract}


\maketitle

At present, laser-nuclear physics usually involves
secondary particles such as electrons in a plasma~\cite{plasma}. 
This indirect technique allows to reach field strengths that
can induce various high-energy processes such as
nuclear fusion and fission or particle acceleration~\cite{fusion}.
On the other hand, especially quantum optics demonstrates
that the direct interaction of laser fields with atoms
enables one to modify or even control the atomic dynamics,
with a multitude of applications~\cite{scully-zubairy,neugart02}.
Thus the question arises, whether direct interactions
with super-intense laser fields could also be
employed in nuclear physics. 
While the coupling of electric and nuclear transitions has been 
studied before~\cite{el-nucl}, direct laser-nucleus 
interactions traditionally have been dismissed.
Mostly, this was based on too small interaction matrix
elements~\cite{matinyan98}. 
Some exceptions are the interaction of x-ray laser fields
with nuclei in relation to $\beta$ decay~\cite{becker83}
and x-ray-driven gamma emission of nuclei \cite{carroll01}.
With the advent of new coherent x-ray laser sources
in the near future, however, these conclusions have to 
be reconsidered.

Therefore in this Letter, we demonstrate that currently envisaged  
high-frequency lasing and ion accelerator 
technology does allow for the direct resonant 
interaction of laser fields with nuclei.
Besides the proof of principle, these interactions may be
utilized e.g. for the optical measurement of nuclear 
properties such as transition frequency and 
dipole moment, thus opening the field of nuclear quantum 
optics. As an explicit example, we show that nuclei 
may be prepared in excited states in a controlled manner 
allowing for the study of nuclear reactions with 
excited nuclei. 
The time evolution of this process allows to extract
nuclear parameters such as transition dipole moments
free of nuclear model assumptions.
We discuss requirements and limitations,
as well as possible observables and applications.
A key advantage of coherent
x-ray laser light is that it, in principle, allows to study
phenomena well known from atomic systems such as
photon echos, coherent trapping or electromagnetic induced 
transparency~\cite{scully-zubairy}.
This depends on the nuclear excitation spectra,
and considerably increases the demands on the employed light source
and target preparation. 
As ultimate goal, one may hope that  direct laser-nucleus
interactions could become a versatile tool to enhance 
preparation, control and detection in nuclear physics.


Nuclei throughout the nuclear chart exhibit various kinds 
of excitations. The most prominent and simple ones in terms of 
theoretical understanding are probably (quadrupole-type) 
vibrations in even-even spherical systems and rotations in 
even-even deformed nuclei. However, depending on the 
nuclear system, quite complicated excitations and couplings 
between them can arise.
Many actinide nuclei possess rather low (collective) E1 
excitations~\cite{Aas}.
These E1 transitions can be found, e.g.,  in alternating 
parity rotating bands. They are related to the collective potential 
of these nuclei and the interplay between quadrupole and octupole 
degrees of freedom in this area of the nuclear chart.
But also other nuclei have E1 transitions with similar
properties, hence the physics described here is not limited to a 
few special cases. Some example transitions are listed in Tab.~\ref{nuclei}.
We focus on transitions
starting from metastable ground states, but transitions between
excited states could be studied as well even though they are
harder to prepare. 


We consider the nucleus as a pure two-level system that can 
be described by the state vector 
$|\psi\rangle = C_g |g\rangle \, + C_e |e\rangle$,
where $|g\rangle$ denotes the nuclear ground state and 
$|e\rangle$ denotes the excited state~\cite{scully-zubairy}. 
This approach is justified by the fact that even though
we consider super-intense laser fields, on a nuclear
scale, the induced perturbation is moderate. This 
allows to neglect  relativistic effects
and interactions beyond the electric dipole approximation,
and to focus on near-resonantly
driven transitions.
The Gaussian laser pulse is given by $E(t) \, \sin(\nu t)$, where 
$E(t)$ is the (time-dependent) electric field amplitude~\cite{bauer}. 
$\hbar \omega$ and $\mu$ are the transition energy and the dipole moment,
and $\nu$ is the frequency of the laser. 
The Rabi frequency is given by $\Omega(t) = \mu E(t)/\hbar$.
%
%
%
\begin{table}[t]
\begin{tabular}{l|c|c|c|c|c}
nucleus	& transition & $\Delta E$ [keV] & $\mu$ [e\,fm]& $\tau(g)$ & $\tau(e)$
[ps]
\\ \hline
$^{153}$Sm  & $3/2^- \! \to \! 3/2^+$ &  35.8  & 
$>0.75^{1)}$  & 47 h & $< 100$ \\
$^{181}$Ta  & $9/2^- \! \to \! 7/2^+$ &  6.2  & 
0.04$^{1)}$  & stable & $6\cdot10^6$ \\
$^{225}$Ac  & $3/2^+ \! \to \! 3/2^-$ &  40.1  & 
0.24$^{1)}$  & 10.0 d & 720 \\
$^{223}$Ra  & $3/2^- \! \to \! 3/2^+$ &  50.1  & 
0.12  & 11.435 d & 730 \\
$^{227}$Th & $3/2^- \! \to \! 1/2^+$ & 37.9  & -$^{2)}$ 
& 18.68 d & -$^{2)}$ \\
$^{231}$Th & $5/2^- \! \to \! 5/2^+$ & 186  & 0.017 
& 25.52 h & 1030 
\end{tabular}
\caption{Parameters of few relevant nuclear systems and 
E1 transitions~\cite{Aas}.
The transitional energy, the dipole moment, 
and the life times of the ground and excited state  are 
denoted by $\Delta E$, $\mu$, $\tau(g)$, and $\tau(e)$, 
respectively. Dipole moments with super-index 1) are estimated
via the Einstein A coefficient from $\tau(e)$ and $\Delta E$;
values with index 2) are not listed in~\cite{Aas}.}
\label{nuclei}
\end{table}
%
The time evolution of the nuclear transition under the
influence of the laser field pulse can conveniently
be described via a master equation treatment for the
system density matrix $\rho$, which allows to consider
additional dephasing rates for the nuclear coherences.
This is required, as most high-frequency laser facilities 
suffer from a limited coherence time even within single
field pulses, in contrast to typical 
low-intensity cw laser systems as utilized in atomic physics.
In a suitable interaction picture, the master equation 
reads ($A_{ij} = |i\rangle\langle j|$ for $i,j\in\{e,g\}$)
\begin{align}
\frac{\partial \rho}{\partial t} = &
\frac{i}{\hbar}[H_0,\rho]
- \frac{\gamma_{SE}}{2} \left ( [A_{eg},A_{ge}\rho] + \textrm{h.c.}
  \right) 
\nonumber \\
& - \gamma_d \left ( [A_{ee},A_{ee}\rho] + \textrm{h.c.} \right ) \,,
\label{master}
\end{align}
where $H_0 = \hbar \Delta A_{ee} + \hbar\Omega(t)
(A_{eg}+A_{ge})/2$ with detuning $\Delta = \nu - \omega$.
The spontaneous emission rate from the upper level
is $\gamma_{SE}$, and $\gamma_d$ is an additional dephasing rate 
to model laser field pulses with limited coherence times.
Purely coherent pulses correspond to $\gamma_d=0$.
We further define the inversion between the two nuclear
levels, given by 
$W(t) = \langle g|\rho |g\rangle - \langle e|\rho | e\rangle
= |C_g|^ 2 - |C_e|^2$.

We have been led by the laser specifications of 
current x-ray laser design reports for TESLA XFEL 
at DESY~\cite{tesla} and 
XRL at GSI~\cite{xrl}, see Tab.~\ref{lasers}.
Acceleration of the target ions allows to
bring the laser in resonance with 
nuclear transitions above the maximum photon energy.
This, however, demands a major 
experimental facility 
that provides both suitable x-ray laser and nuclear beams. 
This step may not be required for 
next-generation laser sources or later stages of extension
of the ones discussed here. In the meantime, low-energetic
transitions such as in $^{181}$Ta or between excited nuclear
states can be studied without accelerating the target.

\begin{table}[t]
\begin{tabular}{l|c|c|c|c}
 &  $\omega_{\rm max}$ [eV]  & I [W/cm$^2$] & ${\mathcal B}_{\textrm res}$ 
 & I$_{\textrm res}$ [W/cm$^2$]\\ \hline
X-1  & 56 & $10^{15}$  & 895 & $8\cdot 10^{20}$\\
X-2  & 90 & $10^{16}$   & 557 & $3\cdot 10^{21}$\\ \hline
T & 12400 & $10^{16} - 10^{20}$ & 4 & $2\cdot 10^{17}- 2\cdot 10^{21}$
 \end{tabular}
\caption{Example laser configurations employed in this study. 
$\omega_{\rm max}$ is the maximum photon energy. The lab frame intensities
$I$ depend on the focussing of the beam.
The parameter sets X-1/X-2 are inspired by the GSI XRL facility,
the set T by SASE 1 of TESLA XFEL at DESY.
${\mathcal B}_{\textrm res}$ is the required factor $(1+\beta)\gamma$
to match the nuclear rest frame laser frequency with the transition frequency 
in  $^{223}$Ra (see. Tab.~\ref{nuclei}). I$_{\textrm res}$ is the
laser intensity in this rest frame.}
\label{lasers}
\end{table}

%
In the following, we work in the 
nuclear rest frame. Thus our treatment
is independent of the particular setup, be it a powerful 
laser source with a resting nucleus, or an accelerated nucleus 
with a less powerful laser beam. 
In the rest frame of the nucleus (subscript $N$), the 
Doppler shifted laboratory frame (subscript $L$) electric 
field strength $E$ and laser frequency $\nu$ are given by
\begin{eqnarray}
E_N &=& \sqrt{(1+\beta)/(1-\beta)} E_L = 
  (1+\beta)\gamma E_L 
\\
\nu_N &=& \sqrt{(1+\beta)/(1-\beta)} \nu_L = 
  (1+\beta)\gamma \nu_L 
\,.
\end{eqnarray}
Table~\ref{lasers} shows the factors
$(1+\beta)\gamma$ required to match rest-frame laser frequency
and the transition frequency of $^{223}$Ra, 
along with the laser intensity in this frame.

Table~\ref{nuclei} lists typical transition data for 
nuclear systems under investigation here~\cite{Aas}. 
Note that the ground states are metastable
in our context, which simplifies 
the preparation and acceleration of the nuclei.
In several cases ($^{153}$Sm, $^{223}$Ra, $^{227}$Th, $^{231}$Th), 
a third level exists between the 
ground state and the dipole-allowed excited state.
Due to a branching ratio of 100:2.6 in $^{223}$Ra, this 
system still is an excellent approximation to a pure
two-level system. But 
in $^{227}$Th, the branching ratio of the E1 excited state 
to the two lower states (at 0 keV and 9.3 keV) is 100:96, thus
forming a three-level system in $\Lambda$-configuration. 
This difference will be discussed below.

\begin{figure}[t]
\centerline{\epsfxsize=8.6cm \epsfbox{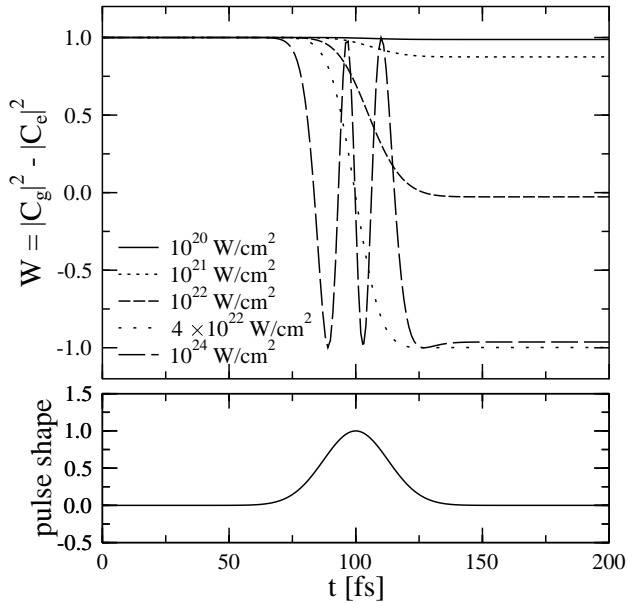}}
\caption{Inversion $W$ (top) and electric field envelope 
of the 30 fs (FWHM) Gaussian laser pulse (bottom) as functions
of time in the nuclear rest frame for the E1 transition 
in $^{223}$Ra. $|C_g|^2$ and $|C_e|^2$ denote the occupation
probabilities of the ground (g) and excited (e) state, respectively.}
\label{30fs_pulse_ra}
\end{figure}

\begin{figure}[t]
\centerline{\epsfxsize=8.6cm \epsfbox{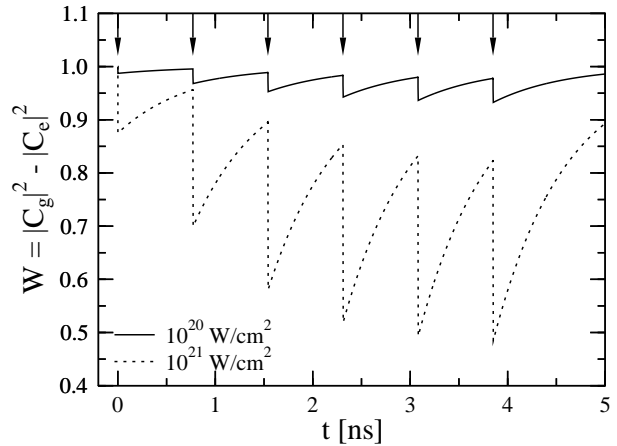}}
\caption{Inversion $W$ as function
of time in the nuclear rest frame for the E1 transition in 
$^{223}$Ra and intensities as indicated. The maxima of 
the train of six 30 fs (FWHM) Gaussian laser pulses are 
indicated by the downward arrows.}
\label{5_pulses}
\end{figure}

We now consider the transition 
in $^{223}$Ra as a typical example which requires 
moderate pre-acceleration of the nuclei. 
Figure~\ref{30fs_pulse_ra} 
displays the inversion of the nuclear E1 transition in 
$^{223}$Ra for a 30 fs (FWHM) 
pulse and various laser intensities 
in the nuclear rest frame. As expected, the dynamics of the
two-level system strongly depends on the laser intensity. 
While for the lowest intensity shown the system remains 
almost in the ground state ($W\approx 1$), with increasing order of the 
intensity it is more affected until it oscillates rapidly 
for $I = 10^{24}$~W/cm$^2$. A $\pi$ pulse that would
directly transfer the system to the excited state without 
further oscillations can be found around
$I_\pi \approx 4\cdot 10^{22}$ W/cm$^2$. 
A series of pulses can further increase the excitation, 
given that the time between the 
pulses is of similar order or smaller than the life time of the 
excited nuclear state. That way, subsequent pulses will enhance the
inversion, which will decrease by a smaller amount in between the
pulses. Figure~\ref{5_pulses} displays such a 
scenario for two different intensities with
a train of 6 pulses. 
The chosen bunch repetition time corresponds
to the fundamental minimum of 770ps given 
in the TESLA design report~\cite{tesla}.
%

\begin{figure}[t]
\centerline{\epsfxsize=8.6cm \epsfbox{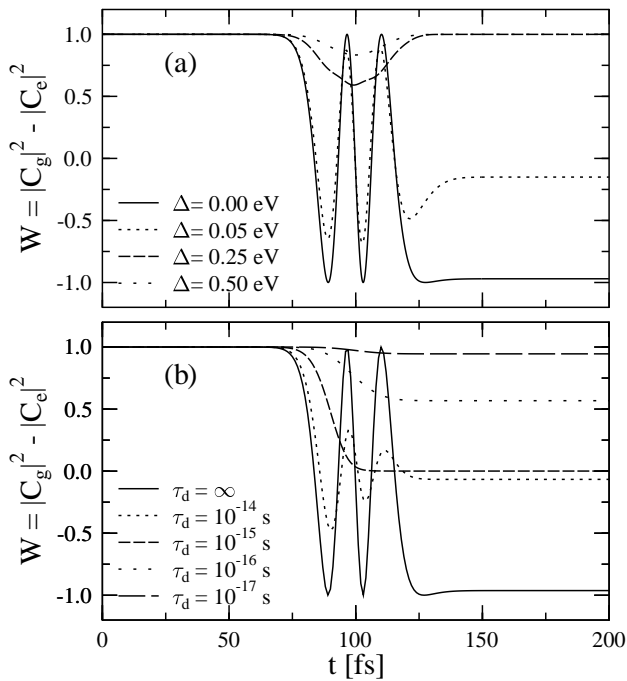}}
\caption{Upper part (a): Inversion $W$ versus
laser field detuning in
$^{223}$Ra for a 30 fs (FWHM) Gaussian laser pulse.
Lower part (b): The inversion for
different decoherence times. Both cases correspond to $I=10^{24}$~W/cm$^2$.}
\label{detuning}
\end{figure}

%
Next, in Fig.~\ref{detuning}(a), the influence of the
laser field detuning is shown. The 
excitation probability depends sensitively on the resonance 
condition, which, however, may be relaxed by the
bandwidth of the laser pulse.

Up to now, we have considered coherent laser field pulses.
In high-frequency laser facilities, however, the coherence
time typically is smaller than the pulse duration. 
We have thus added the additional cross damping rate $\gamma_d$ in 
Eq.~(\ref{master}), which is set to values around the
inverse coherence time. Results are shown in Fig.~\ref{detuning}(b).
The Rabi oscillations are damped stronger with
decreasing coherence length, until the inversion $W$ remains
positive for mostly incoherent light. 
Note that the decrease in the inversion $W$ 
can partially be countered by increasing
the field intensity. Thus it is possible to observe the
partial inversion with largely incoherent fields,
and thus measure nuclear parameters such as the dipole moments, see below.
The limited coherence implies a spectral broadening
of the laser pulse, which is further increased by the finite
energy resolution of the ion accelerator if acceleration 
is required. This decreases the number of resonant photons in
the laser field, and thus leads to a reduction of the signal yield.
Therefore long coherence times and high-quality beams or
fixed targets are desirable, as they enhance
the experimental possibilities. Note that
the TESLA design report contains an extension to a two-stage
FEL which would provide highly coherent light of low bandwidth with
an increase in brilliance of about 500 as compared to the single-stage
FEL considered here~\cite{tesla}.

%
%

For nuclear transitions, the transitional dipole moment 
$\mu$ is usually extracted with the help of the measured 
reduced transition probability 
$B(E1; I_i \rightarrow I_f)$~\cite{Aas}. 
The rotational model formula often used in the extraction 
of the dipole moment $\mu$ is given by
$B(E1; I_i \rightarrow I_f) = 3/(4\pi) \times \mu^ 2 
\times \langle I_i K_i 10 | I_f K_f\rangle^2$. 
This formula involves assumptions on the structure 
of the nucleus, namely that the nucleus is a perfect rotator, 
and that the moment of inertia is identical for the levels
involved.
In contrast, the determination of $\mu$ with the help of x-ray 
lasers  constitutes an optical and more direct alternative. 
Measurements of the response as a function
of the pulse parameters yield an excitation function from which 
the dipole moment can be extracted. This method is free from 
any assumptions on the nuclear structure but the
two-level approximation, which is well justified
(see Fig.~\ref{detuning}(a)). 
At the same time,
the dependence on the detuning could allow to measure the
nuclear transition frequency.
Determination of the dipole matrix element $\mu$ via both
methods provides information about the nucleus structure
and the validity of the nuclear model assumptions. 
This will enrich our knowledge on nuclear structure and 
interactions between the nucleons.

The controlled excitation of nuclei with x-ray laser fields,
or even nuclear Rabi oscillations, can be detected in several ways. 
First, fluorescence radiation is emitted during the process, which could 
be detected as a function of the applied field pulse.
A time discrimination of the detector allows to separate between
the immediate scattering and the spontaneous emission due to 
real excitation of the transition, and thus to avoid the primary
sources of background noise, but requires fast detectors.
%
%
%
%
In contrast to fixed targets and depending on the lifetime of
the excited state, one could also stop and capture
the accelerated nuclei, e.g., using {\em implementation} 
methods~\cite{chemical}, or measure spontaneously
emitted photons behind the interaction region rather than
gating the detectors electronically.
If no target acceleration is needed, then 
a fixed sample can be used,
which may allow to increase the target particle density and thus
the signal yield.
From the design report for SASE 1 at TESLA XFEL and parameters for current and
future ion beam sources~\cite{ionbeam}, the signal rate
due to spontaneous emission after real excitations of the nuclei
can be estimated. For nuclei accelerated with an energy resolution of
0.1\% such that 12.4\:keV photons produced by SASE 1 
become resonant 
with the E1 transition in $^{223}$Ra, the total photon energy
spread in the nuclear rest frame is about 67\:eV. From the
peak photon brilliance 
one may estimate a flux of approx. $4.1 \times 10^{18}$ photons/second 
resonant within the transition width of the excited state.
Assuming in the lab frame a focal diameter of $20\mu$m, focal length of twice
the Rayleigh length, and laser pulse duration of 100fs, 
then the signal photon yield per laser
pulse per single target nucleus is about $5.4\times 10^{-10}$. 
This amounts to a signal of about 1.4 emitted photons 
per day for a single  nucleus. With $2.5\times 10^{10}$ particles in 
a bunch length of $\tau=50$ns in an ion beam of 2mm diameter
as  target~\cite{ionbeam} (particle density $5.3\times 10^8$ cm$^{-3}$), one 
estimates $2.6\times 10^{-4}$ signal photons per 
pulse and $6.8 \times 10^5$ photons per day. 
For a second set of parameters labelled SIS100/FAIR  in~\cite{ionbeam} 
with particle density $10^{11}$cm$^{-3}$, one finds
 $5.3\times 10^{-2}$ signal photons per 
pulse and $1.4 \cdot 10^8$ photons per day. 
Note, however, that the photons per day assume
a matching of ion and laser pulse repetition rate.

A second measurement principle involves nuclear state detection,
which requires a dependence of secondary processes on
the internal state of the nucleus. Similar techniques are
used in atomic physics, if the detection of signal photons e.g. over a
thermal background is difficult~\cite{state-detection}.
The state detection methods could also be possible via
{\em nuclear shelving}, similar to the electron shelving
in atomic physics~\cite{shelving}. For example, the
excited $^{227}$Th nucleus has a high branching
ratio to a second metastable lower level. Thus a
repeated excitation of the nucleus, e.g. in an ion
storage ring, would provide selective optical pumping 
between the two metastable lower states, which could be
detected in a subsequent secondary process without the need
for fast detector gating.

\begin{acknowledgments}
TJB thanks C. M\"uller and U. D. Jentschura for helpful discussions.
\end{acknowledgments}


\end{document}